\documentclass[aps,prd,preprint,superscriptaddress,tightenlines,nofootinbib]{revtex4}

\usepackage{graphicx}
\usepackage{dcolumn}
\usepackage{bm}

\newcommand{\gamgam}{\gamma\gamma}
\newcommand{\jpipi}{\pi^{+}\pi^{-}{\mathrm{J}}/\psi}

\newcommand{\pipillbr}{\pi^{+}\pi^{-}(l^{+}l^{-})}

\newcommand{\deltm}{\mathrm{M}(\pi^{+}\pi^{-}l^{+}l^{-}) - 
		    \mathrm{M}(l^{+}l^{-})}

\newcommand{\jll}{{\mathrm{J}}/\psi \rightarrow l^{+}l^{-}}
\newcommand{\jee}{{\mathrm{J}}/\psi \rightarrow 
                  {\mathrm{e}}^{+}{\mathrm{e}}^{-}}
\newcommand{\jmm}{{\mathrm{J}}/\psi \rightarrow \mu^{+}\mu^{-}}

\newcommand{\twophprod}{{\mathrm{e}}^{+}{\mathrm{e}}^{-} \rightarrow 
			{\mathrm{e}}^{+}{\mathrm{e}}^{-}(\gamgam)}
\newcommand{\isrx}{{\mathrm{e}}^{+}{\mathrm{e}}^{-} \rightarrow 
                   \gamma{\mathrm{X(3872)}}}
\newcommand{\Gammagg}{\Gamma_{\gamgam}({\mathrm{X(3872)}})}
\newcommand{\Gammaee}{\Gamma_{{\mathrm{ee}}}({\mathrm{X(3872)}})}

\newcommand{\resDecayBR}{{\cal{B}}(\mathrm{R} \rightarrow \jpipi)}

\newcommand{\xBR}{{\cal{B}}(\mathrm{X(3872)} \rightarrow \jpipi)}

\newcommand{\xGGUL}{(2\mathrm{J}+1)\Gammagg\xBR}
\newcommand{\xISRUL}{\Gammaee\xBR}

\begin{document}

\preprint{CLEO CONF 04-07}         
\preprint{ICHEP04 ABS10-0768}      

\title{Search for X(3872) in Untagged $\gamgam$ Fusion and Initial State 
Radiation Production with CLEO III}
\thanks{Submitted to the 32$^{\rm nd}$ International Conference on High 
Energy Physics, Aug 2004, Beijing}


\author{Z.~Metreveli}
\author{K.~K.~Seth}
\author{A.~Tomaradze}
\author{P.~Zweber}
\affiliation{Northwestern University, Evanston, Illinois 60208}
\author{J.~Ernst}
\author{A.~H.~Mahmood}
\affiliation{State University of New York at Albany, Albany, New York 12222}
\author{K.~Arms}
\author{K.~K.~Gan}
\affiliation{Ohio State University, Columbus, Ohio 43210}
\author{D.~M.~Asner}
\author{S.~A.~Dytman}
\author{S.~Mehrabyan}
\author{J.~A.~Mueller}
\author{V.~Savinov}
\affiliation{University of Pittsburgh, Pittsburgh, Pennsylvania 15260}
% ...
\collaboration{CLEO Collaboration} 
\noaffiliation


\date{July 31, 2004}

\begin{abstract} 
We report on an exclusive search for the X(3872) state in 
the decay $\jpipi$, $\jll$ ($l$ = e,$\mu$) from untagged $\gamgam$ 
fusion and initial state radiation production using a 15.1 fb$^{-1}$ data 
sample with the CLEO III detector.  By taking advantage of the unique 
$\jpipi$ decay kinematics, separate measurements for each production 
process are obtained.  No signals are observed and preliminary upper 
limits have been established as 
$\xGGUL$ $<$ 12.9 eV and $\xISRUL$ $<$ 8.0 eV, 
both at a 90$\%$ confidence level.
\end{abstract}

\maketitle

The Belle Collaboration reported the observation of a narrow state,
X(3872), in the decay of 152 million B$\overline{\mathrm{B}}$ events through 
B$^{\pm}$ $\rightarrow$ K$^{\pm}$ X, X $\rightarrow$ $\jpipi$, $\jll$ 
($l$ = e,$\mu$) measuring M(X) = 3872.0 $\pm$ 0.6 (stat) $\pm$ 0.5 
(syst) MeV/c$^{2}$ and $\Gamma$ $<$ 2.3 MeV/c$^{2}$ 
at a 90$\%$ confidence level (C.L.) \cite{xBELLE}.  The observation 
was confirmed by the CDF II Collaboration \cite{xCDFII} 
and D{\O} Collaboration \cite{xD0} in inclusive production 
with X(3872) decaying to $\jpipi$, $\jmm$ in 
p$\overline{\mathrm{p}}$ collisions at $\sqrt{\mathrm{s}}$ = 1.96 TeV 
and also by the {\slshape{B{\scriptsize{A}}B{\scriptsize{AR}}}} 
Collaboration \cite{xBABAR} 
 using 117 million B$\overline{\mathrm{B}}$ events with the same 
decay channels as the Belle observation.  
CDF II, D{\O}, and {\slshape{B{\scriptsize{A}}B{\scriptsize{AR}}}} 
measured masses of
3871.3 $\pm$ 0.7 (stat) $\pm$ 0.4 (syst) MeV/c$^{2}$, 
3871.8 $\pm$ 3.1 (stat) $\pm$ 3.0 (syst) MeV/c$^{2,}$,
and 3873.4 $\pm$ 1.4 (stat) MeV/c$^{2}$, respectively, and widths consistent
with their respective detector resolutions.

A large number of theoretical papers now exist 
\cite{xccbarbound1,xccbarbound2,xccbartwophot,xmole1,xmole2,xthref1,xthref2,xthref3,xthref4,xthref5,xthref7,xglueball,xdiagdecay} 
with different interpretations of the nature of the X(3872) state 
and its possible quantum numbers.  
Among the possibilities are: (a) a charmonium state; (b) a 
loosely bound ``molecular'' state; 
or (c) an exotic state.  Barnes and Godfrey \cite{xccbarbound1} and Eichten, 
Lane, and Quigg \cite{xccbarbound2} have
examined the charmonium options in detail and conclude that, because of the 
small width of X(3872) and despite the larger predicted masses of the 
candidate charmonium 
states, the viable charmonium options are the negative C parity states of 
1$^{3}$D$_{2}$ (J$^{\mathrm{PC}}$ = 2$^{--}$), 1$^{3}$D$_{3}$ (3$^{--}$), and
2$^{1}$P$_{1}$ (1$^{+-}$), and positive C states of 1$^{1}$D$_{2}$ (2$^{-+}$)
and 2$^{3}$P$_{1}$ (1$^{++}$).  
Tornquist \cite{xmole1} and Swanson 
\cite{xmole2} propose that, since X(3872) is close to the 
D$\overline{\mathrm{D}}$$^{*}$ threshold 
(M(D$^{0}$) + M($\overline{\mathrm{D}}$$^{0*}$) = 3871.3 $\pm$ 0.5
MeV/c$^{2}$\cite{2004partlist}) and there are no published studies 
in $\pi^{0}\pi^{0}$J/$\psi$ decays, the observed decay $\jpipi$ 
proceeds through $\rho^{0}$J/$\psi$, 
$\rho^{0}$ $\rightarrow$ $\pi^{+}\pi^{-}$.  They suggest 
0$^{-+}$ and 1$^{++}$ as the most likely assignment in a molecular model.  
Finally, we note that Morningstar and Peardon \cite{xglueball} predict a
three gluon 1$^{--}$ glueball with a mass of 3850 MeV/c$^{2}$.

Searches for X(3872) have been made but no positive signals have been 
observed in the decay channels X(3872) 
$\rightarrow$ $\gamma\chi_{\mathrm{c1}}$, $\gamma\chi_{\mathrm{c2}}$, 
$\gamma$J/$\psi$, $\eta$J/$\psi$, D$^{+}$D$^{-}$, 
D$^{0}\overline{\mathrm{D}}$$^{0}$, and 
D$^{0}\overline{\mathrm{D}}$$^{0}\pi^{0}$ 
\cite{xBELLE,xCharmPossBELLE,xDDBELLE,xJpsiEtaBABAR}.  
C. Z. Yuan, X. H. Mo, and P. Wang determined an upper 
limit for initial state radiation production of 
$\xISRUL$ $<$ 10 eV (90$\%$ C.L.) \cite{xISRBES}
using a previously published 22.3 pb$^{-1}$ data sample  
of e$^{+}$e$^{-}$ annihilations at $\sqrt{\mathrm{s}}$ = 4.03 GeV with 
the BES detector \cite{BESISR}.
With all of these searches, the new state X(3872) has only been 
experimentally observed decaying to $\pi^{+}\pi^{-}$J/$\psi$. 

The variety of possibilities for the structure of X(3872) suggests that,
irrespective of the models, it is useful to limit the J$^{\mathrm{PC}}$ of
X(3872) as much as possible.  Untagged $\gamgam$ fusion, which only 
populates states of positive C parity and even values of total angular 
momentum, can be used to shed light on the charge parity and 
total angular momentum of X(3872) and its population via initial state 
radiation can determine its leptonic width and its likely vector character.  
The present investigation is designed to study the possible production 
of X(3872) in untagged $\gamgam$ fusion and initial state radiation in 
the final state $\jpipi$, $\jll$ ($l$ = e,$\mu$).

Untagged $\gamgam$ fusion and initial state radiation (ISR) resonance 
production have similar event characteristics.  Untagged $\gamgam$ fusion 
resonance production occurs when a photon is emitted by each incident 
electron/positron, the two almost real photons ``fuse'' to form the resonance, 
while the scattered electron/positron are not detected.  
ISR resonance production predominately occurs when either the initial 
electron or positron emits a hard photon which lowers the 
center-of-mass energy of the electron-positron system to the invariant mass 
of the produced resonance before annihilation.  The radiated photon also 
has an angular distribution which is very sharply peaked along the beam 
axis and therefore is rarely detected.
Both types of resonance production processes have a total observed 
energy (E$_{\mathrm{total}}$) much smaller than the center-of-mass energy of 
the electron-positron system and a small observed transverse momentum.

The data used for this X(3872) search were collected with the detector in 
the CLEO III configuration \cite{CLEOIIIDetector}  
at the Cornell Electron Storage Ring (CESR).
The CLEO III detector is a cylindrically symmetric detector designed to 
study e$^{+}$e$^{-}$ annihilations and provide 93$\%$ coverage of solid 
angle for charged and neutral particle 
identification.  The detector components important for this analysis are 
the drift chamber (DR), CsI crystal calorimeter (CC), and muon identification 
system (MIS). The DR and CC are operated within a 1.5 T magnetic field 
produced 
by a superconducting solenoid located directly outside of the CC.   
The DR detects charged particles and measures their momenta and ionization 
energy loss (dE/dx).  The CC allows precision measurements of 
electromagnetic shower energy and position.  The MIS consists of proportional 
chambers placed between layers of the magnetic field return iron to detect 
charged particles which penetrate a minimum of three nuclear 
interaction lengths.

The data consist of a 15.1 fb$^{-1}$ sample at or near the energies of 
the $\Upsilon(nS)$ resonances, where $n=1$--$5$, and in the vicinity of the 
$\Lambda_{\mathrm{b}}\overline{\Lambda}_{\mathrm{b}}$ threshold.  
Table 1 lists the six different initial center-of-mass energies (CME) and 
integrated luminosities of the data sample.  

\begin{table}[ht]
\caption{Data sample used for this X(3872) search.  $\langle\sqrt{\mathrm{s}}$ 
$\rangle$ is the integrated luminosity averaged initial CME 
for the $\Upsilon$(1S-5S) resonance and threshold 
$\Lambda_{\mathrm{b}}\overline{\Lambda}_{\mathrm{b}}$ data samples, 
respectively, and $\cal{L}$(e$^{+}$e$^{-}$) is the e$^{+}$e$^{-}$ 
integrated luminosity for each CME region.}
\begin{center}
\begin{tabular}{|c|c|c|}
\hline
& $\langle\sqrt{\mathrm{s}}$ $\rangle$ (GeV) & 
$\cal{L}$(e$^{+}$e$^{-}$) (fb$^{-1}$) \\
\hline
$\Upsilon$(1S) & 9.458  & 1.47 \\
$\Upsilon$(2S) & 10.018 & 1.84 \\
$\Upsilon$(3S) & 10.356 & 1.67 \\
$\Upsilon$(4S) & 10.566 & 8.97 \\
$\Upsilon$(5S) & 10.868 & 0.43 \\
$\Lambda_{\mathrm{b}}\overline{\Lambda}_{\mathrm{b}}$ threshold 
& 11.296 & 0.72 \\
\hline
\end{tabular}
\end{center}
\end{table}

The expected characteristics for $\gamgam$ fusion and ISR resonance 
production are studied by generating signal Monte Carlo (MC) 
samples. The $\gamgam$ fusion resonance signal MC process is $\twophprod$, 
$\gamgam$ $\rightarrow$ X(3872) using the formalism of 
Budnev {\itshape{et al.}} \cite{ggcs}. 
The ISR signal MC process is $\isrx$ using the angular distribution 
expression from M. Benayoun {\itshape{et al.}} \cite{isrprod}. 
The X(3872) and J/$\psi$ are decayed 
according to phase space.  The MC simulation 
of the CLEO III detector was based upon GEANT 3.21/11 \cite{GEANTMC} with 
simulated events processed in the same manner as the data.  

A fully reconstructed event has four charged particles and zero net 
charge.  All charged particles must lie within the DR volume and 
satisfy standard requirements for track quality and distance of closest 
approach to the interaction point (IP).  Events must also have detected 
E$_{\mathrm{total}}$ $<$ 6 GeV.  
Signal-to-background studies are performed to 
optimize signal efficiency and background suppression 
using the $\gamgam$ fusion signal MC sample for the signal and background 
from data 
with $\deltm$ above and below the X(3872) search region, i.e., $\deltm$ 
= 0.63--0.7 or 0.85--0.92 GeV/c$^{2}$.  The selection variables optimized are 
the total neutral energy (E$_{\mathrm{neu}}$) of the event, 
total transverse momentum of the four charged tracks (p$_{\mathrm{tr}}$), 
lepton pair invariant mass (M($l^{+}l^{-}$)) of the 
$\jll$ decay, and particle identification (PID) of the charged tracks. 
Based on the optimization studies, events are selected 
with E$_{\mathrm{neu}}$ $<$ 0.4 GeV and p$_{\mathrm{tr}}$ $<$ 0.3 GeV/c.
Events with a $\jee$ decay have both electron candidates satisfying dE/dx 
and shower energy criteria consistent with the electron hypothesis and 
an invariant mass M(e$^{+}$e$^{-}$) = 2.96-3.125 GeV/c$^{2}$.  
Events with a $\jmm$ decay have both muon candidates appearing as a minimum 
ionizing particle in the CC, at least one muon must penetrate a number of 
interaction lengths in the MIS consistent with its momentum, and an invariant 
mass M($\mu^{+}\mu^{-}$) = 3.05-3.125 GeV/c$^{2}$.  Each of the two pions 
recoiling against the J/$\psi$ needs to individually satisfy the 
dE/dx pion hypothesis.

\begin{figure}
\includegraphics*[width=3.75in]{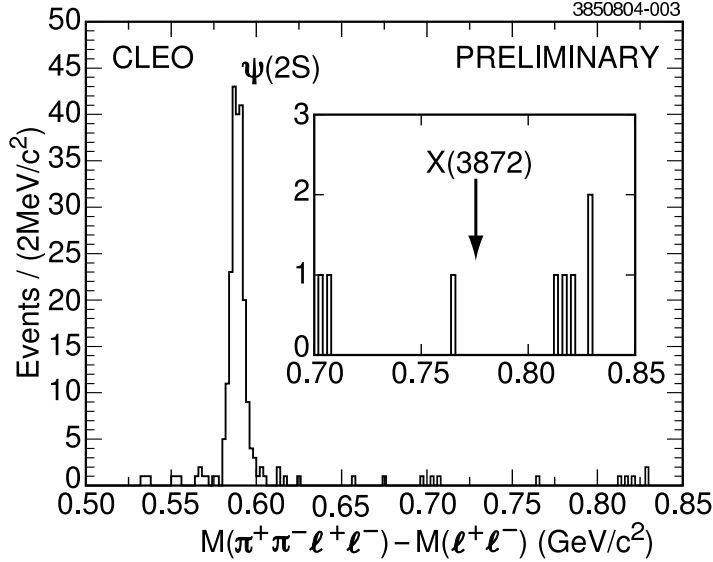}
\caption{$\Delta$M $\equiv$ M($\pi^{+}\pi^{-}l^{+}l^{-}$) - M($l^{+}l^{-}$) 
data distribution for $\Delta$M = 0.514--0.850 GeV/c$^{2}$.  The $\psi$(2S) 
is clearly visible and there is no enhancement in the X(3872) region.}
\end{figure}

Fig. 1 shows the $\Delta$M $\equiv$ M($\pi^{+}\pi^{-}l^{+}l^{-}$) - 
M($l^{+}l^{-}$) data distributions for events which pass the selection 
criteria and have $\Delta$M = 0.514--0.850 GeV/c$^{2}$.  A $\psi$(2S) signal 
is clearly visible while no enhancement is apparent in the X(3872) region.  
A MC sample of $\psi$(2S) produced via ISR is generated to determine the 
efficiency for detecting the observed $\psi$(2S) signal, from which an 
expected number of $\psi$(2S) $\rightarrow$ $\jpipi$, $\pipillbr$ events 
is derived to be 226 $\pm$ 11 events, where the error is from the 
MC efficiency and $\jll$ PDG branching fraction \cite{2002partlist} 
uncertainties.  
The observed number of $\psi$(2S) events is determined by fitting 
the $\psi$(2S) region with the detector resolution and a mass-independent 
background.  The detector resolution is 
determined by fitting the $\psi$(2S) ISR MC resolution function with a 
double Gaussian.  The narrow Gaussian width and relative area and width 
of the double Gaussian are fixed when the $\psi$(2S) is fitted.  The 
observed number of $\psi$(2S) is 206 $\pm$ 15 events and consistent 
with expectations. 

A feature unique to ISR production of charmonium-like resonances which 
decay via $\jpipi$ at initial energies of this data sample is the 
correlation between the tracks of the two leptons.
Fig. 2 shows the two-dimensional cos($\theta$) distributions 
for lepton tracks in the X(3872) ISR resonance and  $\gamgam$ fusion signal 
MC samples.  
A parabolic cut is applied to the two-dimensional lepton pair 
cos($\theta$) distribution to separate the two phenomena 
and is shown in Fig. 2.  
This separation removes $\sim$14$\%$ of the X(3872) $\gamgam$ MC
fusion signal events but rejects more than 99.5$\%$ of the 
ISR events.

\begin{figure}
\includegraphics*[width=4.5in]{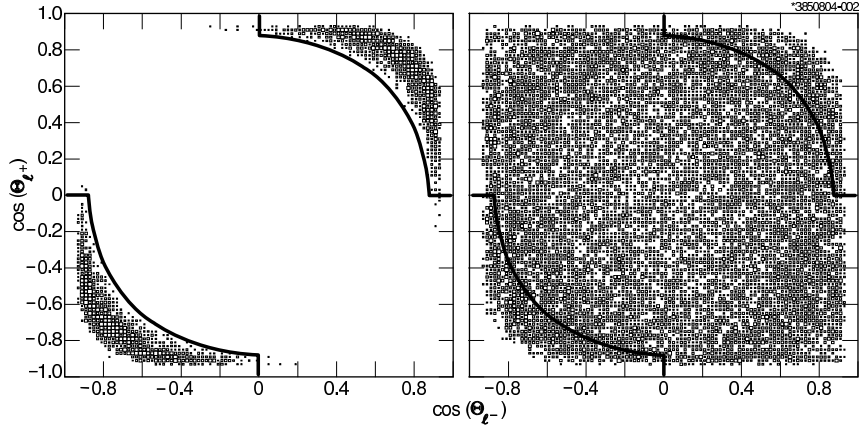} 
\caption{Two dimensional lepton pair cos($\theta$) distribution for the 
X(3872) ISR (left) and $\gamgam$ fusion (right) resonance MC samples.  The 
lines indicate where the ISR resonance and $\gamgam$ fusion samples 
were separated.}
\end{figure}

The total efficiencies in the six different initial CME regions are listed 
in Table 2.  The efficiencies 
are determined from the X(3872) $\gamgam$ fusion and ISR resonance signal 
MC samples.  The same selection criteria is applied to both MC samples 
except with their appropriate two-dimensional lepton pair cos($\theta$) 
correlation.

\begin{table}[ht]
\caption{Detection efficiencies determined from the signal $\gamgam$ fusion 
and ISR resonance MC samples.  $\epsilon_{\mathrm{ee}}$ ($\epsilon_{\mu\mu}$) 
is the total efficiency for detecting X(3872) in events with a $\jpipi$, 
$\jee$ ($\jmm$) decay.  The appropriate $\gamgam$ fusion/ISR separation 
is applied to the respective signal MC sample.}
\begin{center}
\begin{tabular}{|c|c||c|c||c|c|} 
\hline
& & \multicolumn{2}{|c||}{$\gamgam$ Fusion} 
& \multicolumn{2}{|c|}{ISR}\\
\hline
& $\langle\sqrt{\mathrm{s}}$ $\rangle$ (GeV) &
$\epsilon_{\mathrm{ee}}$ & $\epsilon_{\mu\mu}$ &
$\epsilon_{\mathrm{ee}}$ & $\epsilon_{\mu\mu}$ \\
\hline
$\Upsilon$(1S) & 9.458  & 0.128(4) & 0.160(4) & 0.065(3) & 0.083(3) \\
$\Upsilon$(2S) & 10.018 & 0.121(3) & 0.151(4) & 0.054(2) & 0.062(3) \\
$\Upsilon$(3S) & 10.356 & 0.115(3) & 0.137(4) & 0.042(2) & 0.043(2) \\
$\Upsilon$(4S) & 10.566 & 0.123(4) & 0.145(4) & 0.0186(14) & 0.0165(13) \\
$\Upsilon$(5S) & 10.868 & 0.113(3) & 0.139(4) & 0.0025(5) & 0 \\
$\Lambda_{\mathrm{b}}\overline{\Lambda}_{\mathrm{b}}$ threshold & 
11.296 & 0.104(3) & 0.126(4) & 0.0001(1) & 0 \\
\hline
\end{tabular}
\end{center}
\end{table}

The $\Delta$M data distributions in the X(3872) search region for 
$\gamgam$ fusion and ISR resonance production are shown in Fig. 3.
The number of observed events (N$_{\mathrm{obs}}$) is determined by a 
maximum likelihood fit of the $\Delta$M data distribution using a 
production-mode-dependent detector resolution and 
mass-independent background.  The respective detector resolutions 
are determined by a double Gaussian fit of the signal MC resolution 
functions and their shapes are illustrated in Fig. 3.  
For each resolution function, the mean is fixed to the Belle observation 
of $\Delta$M = 0.7751 GeV/c$^{2}$ with the narrow Gaussian width and the 
double Gaussian relative area and width also fixed.  
The upper limits on the observed number of X(3872) events in 
untagged $\gamgam$ fusion and ISR resonance production are 
N$_{\mathrm{obs}}$ $<$ 2.36 and N$_{\mathrm{obs}}$ $<$ 2.1 at 90$\%$ 
C.L., respectively.

\begin{figure}
\includegraphics*[width=3.65in]{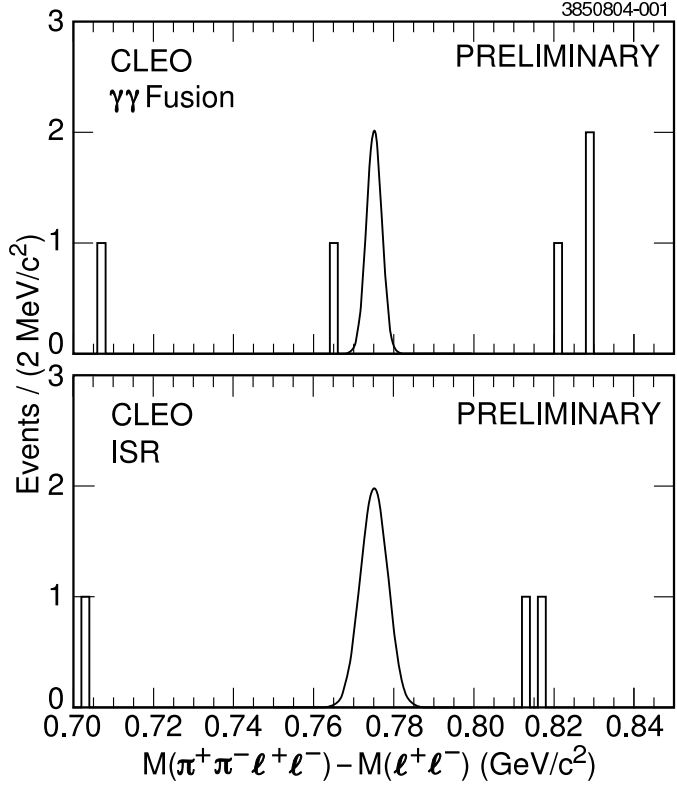}
\caption{$\gamgam$ fusion (top) and ISR (bottom) $\Delta$M 
data distributions with $\Delta$M = 0.7--0.85 GeV/c$^{2}$ satisfying 
respective selection criteria.  The corresponding MC-determined 
resolution functions are illustrated with the means 
fixed at $\Delta$M = 0.7751 GeV/c$^{2}$.}
\end{figure}

The X(3872) production results need to incorporate the six 
different CME regions of the data sample.  The cross section for a 
$\gamgam$ fusion or ISR produced resonance of mass m$_{\mathrm{R}}$ 
decaying to $\jpipi$, $\jll$ is
\begin{equation}
\sigma(\mathrm{R})_{\gamgam,\mathrm{ISR}}\resDecayBR = 
\frac{\mathrm{N_{obs}}}{\sum_{i} \mathrm{f}_{i} 
[\epsilon_{ee,i}{\cal{B}}(\jee)+\epsilon_{\mu\mu,i}{\cal{B}}(\jmm)]}
\end{equation}
where f$_{i}$ = $\mathcal{L}_{i}$(e$^{+}$e$^{-}$)$\sigma$(s$_{i}$,m$_{\mathrm{R}}$)$_{\gamgam,\mathrm{ISR}}$ is the flux for $\gamgam$ 
fusion or ISR production, $\mathcal{L}_{i}$(e$^{+}$e$^{-}$) is the 
e$^{+}$e$^{-}$ integrated luminosity, and $\epsilon_{ee,i}$ 
($\epsilon_{\mu\mu,i}$) is the detection efficiency for events 
with $\jpipi$, $\jee$ ($\jmm$) decays at an initial CME, 
$\sqrt{\mathrm{s}_{i}}$.  $\resDecayBR$ is the R $\rightarrow$ $\jpipi$ 
branching fraction while ${\cal{B}}(\jee)$ and ${\cal{B}}(\jmm)$ 
are the PDG branching fractions \cite{2002partlist}. 
For a narrow resonance, 
$\sigma$(s$_{i}$,m$_{\mathrm{R}}$)$_{\gamgam}$ is defined as \cite{ggcs}
\begin{equation}
\sigma(\mathrm{s}_{i},\mathrm{m_{R}})_{\gamgam} = 
(\frac{\alpha}{\pi\mathrm{m_{R}}})^{2}
~[\mathrm{f}(\frac{\mathrm{m}^{2}_{\mathrm{R}}}{\mathrm{s}_{i}})
(\ln(\frac{\mathrm{s}_{i}}{\mathrm{m}^{2}_{\mathrm{e}}})-1)^{2} 
- \frac{1}{3}(\ln(\frac{\mathrm{s}_{i}}{\mathrm{m}^{2}_{\mathrm{R}}}))^{3}],
\end{equation}
\begin{displaymath}
\mathrm{f}(\frac{\mathrm{m}^{2}_{\mathrm{R}}}{\mathrm{s}_{i}}) = 
(1 + \frac{\mathrm{\mathrm{m}^{2}_{\mathrm{R}}}}{2\mathrm{s}_{i}})^{2} 
\ln(\frac{\mathrm{s}_{i}}{\mathrm{m}^{2}_{\mathrm{R}}}) - \frac{1}{2}
(1 - \frac{\mathrm{m}^{2}_{\mathrm{R}}}{\mathrm{s}_{i}})
(3 + \frac{\mathrm{m}^{2}_{\mathrm{R}}}{\mathrm{s}_{i}})
\end{displaymath}
and $\sigma$(s$_{i}$,m$_{\mathrm{R}}$)$_{\mathrm{ISR}}$ is \cite{isrprod}
\begin{equation}
\sigma(\mathrm{s}_{i},\mathrm{m_{R}})_{\mathrm{ISR}} = 
\frac{2\alpha}{\pi\mathrm{x}_{i}\mathrm{s}_{i}}
(2\ln\frac{\sqrt{{\mathrm{s}_{i}}}}
{\mathrm{m_{e}}}-1)(1-\mathrm{x}_{i}+\frac{\mathrm{x}_{i}^{2}}{2})
\end{equation}
where x$_{i}$ is the ratio of the ISR photon energy to the initial beam 
energy ($\frac{2{\mathrm{E}_{\gamma,i}}}{\sqrt{{\mathrm{s}}_{i}}}$) and 
$\mathrm{m_{e}}$ is the electron mass.
$\sigma$(R)$_{\gamgam}$ is related to the $\gamgam$ fusion 
resonance production by 
\begin{equation}
\sigma(\mathrm{R})_{\gamgam} = 
\frac{8\pi^{2}\mathrm{(2J+1)}\Gamma_{\gamgam}\mathrm{(R)}}
{\mathrm{m}_{\mathrm{R}}}
\end{equation}
and $\sigma$(R)$_{\mathrm{ISR}}$ is related to the ISR 
resonance production by 
\begin{equation}
\sigma(\mathrm{R})_{\mathrm{ISR}} = 
\frac{12\pi^{2}\Gamma_{\mathrm{ee}}\mathrm{(R)}}
{\mathrm{m}_{\mathrm{R}}}
\end{equation}
where $\Gamma_{\gamgam}$(R) ($\Gamma_{\mathrm{ee}}$(R)) is the 
two-photon (e$^{+}$e$^{-}$) partial width of the resonance.  

The sources of systematic uncertainty arise from possible biases in the 
detection efficiency and estimated background level.  
These are studied by varying the track quality, IP, $\gamgam$ 
fusion/ISR separation, and selection criterion optimized in the 
signal-to-background studies.  
Other systematic uncertainties are 
from the e$^{+}$e$^{-}$ luminosity measurement and PDG $\jll$ branching 
fractions.  The total systematic uncertainty contributing to $\gamgam$ 
fusion (ISR) resonance production is 18.5$\%$ (23.4$\%$), determined by 
summing individual contributions in quadrature, and is incorporated into 
the final results by increasing the measured upper limits by 
the respective systematic uncertainties.  The specific X(3872) angular 
distributions for a given J$^{\mathrm{PC}}$ have not been considered in 
this analysis.

Untagged $\gamgam$ fusion resonance production tests the possibility 
of a state having positive C parity and an even value of total 
angular momentum.  The preliminary upper limit for X(3872) $\gamgam$ 
fusion production is 
\begin{displaymath}
\xGGUL < 12.9~\mathrm{eV} 
\end{displaymath}
at a 90$\%$ C.L.  
Assuming $\cal{B}$(B$^{\pm}$ $\rightarrow$ K$^{\pm}$X(3872)) $\approx$ 
$\cal{B}$(B$^{\pm}$ $\rightarrow$ K$^{\pm}\psi$(2S)) = 
(6.8$\pm$0.4)$\times$10$^{-4}$ \cite{2004partlist} leads to $\xBR$ 
$\approx$ 0.02 for both the Belle 
\cite{xBELLE} and {\slshape{B{\scriptsize{A}}B{\scriptsize{AR}}}} 
\cite{xBABAR} results.  This translates into 
(2J+1)$\Gamma_{\gamgam}$(X(3872)) $<$ 0.645 keV (90$\%$ C.L.) for this 
analysis.  The X(3872) (2J+1)$\Gamma_{\gamgam}$(R) is more than four times 
smaller than for $\eta_{\mathrm{c}}$, $\chi_{c0}$, and $\chi_{c2}$, 
i.e., (2J+1)$\Gamma_{\gamgam}$($\eta_{\mathrm{c}}$) = 7.4 $\pm$ 2.3 keV, 
(2J+1)$\Gamma_{\gamgam}$($\chi_{c0}$) = 2.6 $\pm$ 0.5 keV, and 
(2J+1)$\Gamma_{\gamgam}$($\chi_{c2}$) = 2.6 $\pm$ 0.3 keV 
\cite{2004partlist}, respectively.  If X(3872) is a vector meson, the 
preliminary 90$\%$ C.L. upper limit for ISR resonance production is 
\begin{displaymath}
\xISRUL < 8.0~\mathrm{eV}.
\end{displaymath} 
From the observed $\psi$(2S) signal, the upper limit for the X(3872) 
ISR production yield is N$_{\mathrm{X(3872)}}$/N$_{\psi\mathrm{(2S)}}$ 
$<$ 0.011 (90$\%$ C.L.).  The ISR production 
upper limit from this X(3872) search includes our estimated systematic 
uncertainty and is comparable to the ISR 
upper limit of $\xISRUL$ $<$ 10 eV (90$\%$ C.L.) using the BES 
data \cite{xISRBES}.

We gratefully acknowledge the effort of the CESR staff in providing us 
with excellent luminosity and running conditions.  M. Selen thanks the 
Research Corporation, and A.H. Mahmood thanks the Texas Advanced 
Research Program.  This work was supported by the National Science 
Foundation and the U.S. Department of Energy.






\begin{thebibliography}{99}

\bibitem{xBELLE}Belle Collaboration, S. K. Choi {\itshape{et al.}},  
	Phys. Rev. Lett. {\bf 91}, 262001 (2003). 
\bibitem{xCDFII}CDF II Collaboration, D. Acosta {\itshape{et al.}},  
	hep-ex/0312021, Phys. Rev. Lett. (submitted).
\bibitem{xD0} D{\O} Collaboration, V. M. Abazov  {\itshape{et al.}}, 
	hep-ex/0405004, Phys. Rev. Lett. (submitted).  The X(3872) mass 
	was published as M($\pi^{+}\pi^{-}\mu^{+}\mu^{-}$) - 
	M($\mu^{+}\mu^{-}$) = 774.9 $\pm$ 3.1 (stat) $\pm$ 3.0 (syst) 
	MeV/c$^{2}$ and it is converted here to the X(3872) mass assuming 
	the PDG J/$\psi$ mass \cite{2002partlist}.
\bibitem{xBABAR}{\slshape{B{\scriptsize{A}}B{\scriptsize{AR}}}} 
	Collaboration, B. Aubert 
	{\itshape{et al.}}, hep-ex/0406022, Phys. Rev. Lett. (submitted).
\bibitem{2002partlist}Review of Particle Properties, 
	 K. Hagiwara {\itshape{et al.}}, Phys. Rev. {\bf D66}, 010001 (2002).
\bibitem{xccbarbound1}Ted Barnes and Stephen Godfrey, 
	Phys. Rev. {\bf D69}, 054008 (2004).  
\bibitem{xccbarbound2}Estai J. Eichten, Kenneth Lane, and Chris Quigg, 
	Phys. Rev. Lett. {\bf 89}, 162002 (2002); 
	Estai J. Eichten, Kenneth Lane, and Chris Quigg, 
	Phys. Rev. {\bf D69}, 094019 (2004). 
\bibitem{xccbartwophot}E. S. Ackleh and T. Barnes, 
	Phys. Rev. {\bf D45}, 232 (1992).
\bibitem{xmole1}Nils A. T$\ddot{\mathrm{o}}$rnqvist, 
	Phys. Lett. {\bf B590}, 209 (2004).
\bibitem{xmole2}Eric S. Swanson, Phys. Lett. {\bf B588}, 189 (2004).
\bibitem{xthref1}Frank E. Close and Philip R. Page, 
	Phys. Lett. {\bf B578}, 119 (2004).
\bibitem{xthref2}Sandip Pakvasa and Mahiko Suzuki, 
	Phys. Lett. {\bf B579}, 67 (2004).
\bibitem{xthref3}M. B. Voloshin, Phys. Lett. {\bf B579}, 316 (2004).
\bibitem{xthref4}Cheuk-Yin Wong, Phys. Rev. {\bf C69}, 055202 (2004).
\bibitem{xthref5}Eric Braaten and Masaoki Kusunoki, 
	Phys. Rev. {\bf D69}, 074005 (2004).
\bibitem{xthref7}P. Bicudo, hep-ph/0401106.
\bibitem{xglueball}Colin J. Morningstar and Mike Peardon, 
	Phys. Rev. {\bf D60}, 034509 (1999).
\bibitem{xdiagdecay}Eric S. Swanson, hep-ph/0406080.
\bibitem{2004partlist}Review of Particle Properties, 
	S. Eidelman {\itshape{et al.}}, Phys. Lett. {\bf B592} , 1 (2004).
\bibitem{xCharmPossBELLE}S. K. Choi, For the Belle Collaboration, Presented 
	at Lake Louise Winter Institute 2004, 15-21 Feb., Alberta, Canada. 
\bibitem{xJpsiEtaBABAR}{\slshape{B{\scriptsize{A}}B{\scriptsize{AR}}}} 
	Collaboration, B. Aubert {\itshape{et al.}}, 
	Phys. Rev. Lett. {\bf 93}, 041801 (2004).
\bibitem{xDDBELLE} Belle Collaboration, R. Chistov {\itshape{et al.}}, 
	 Phys. Rev. Lett. {\bf 93}, 051803 (2004).
\bibitem{xISRBES}C. Z. Yuan, X. H. Mo, P. Wang, 
	Phys. Lett. {\bf B579}, 74 (2004).
\bibitem{BESISR}BES Collaboration, J. Z. Bai {\itshape{et al.}}
	Phys. Rev. {\bf D57}, 3854 (1998).
\bibitem{CLEOIIIDetector}
Y.~Kubota {\it et~al.}, {Nucl. Instrum. Meth.}, \textbf{A320}, {66} ({1992});
G.~Viehhauser {\it et~al.}, {Nucl. Instrum. Meth}. \textbf{A462}, 
{146} ({2001}); D. Peterson {\it et~al.}, {Nucl. Instrum. Meth.}, 
{\bf A478}, {142} ({2002}); M. Artuso {\it et~al.}, {Nucl. Instrum. Meth.}, 
{\bf A502}, {91} ({2002}).
\bibitem{GEANTMC}R. Brun {\itshape{et al.}}, 
	CERN Long Writeup W5013 (1994).
\bibitem{ggcs}V. M. Budnev, I. F. Ginzburg, G. V. Meledin, and V. G. Serbo, 
	Phys. Reports {\bf 15C}, 181 (1975). 
\bibitem{isrprod}M. Benayoun, S. I. Eidelman, V. N. Ivanchenko, 
	Z. K. Silagadze, Mod. Phys. Lett. {\bf A14}, 2605 (1999). 

\end{thebibliography}
\end{document}